\documentclass[fleqn,10pt]{wlscirep}
\usepackage[utf8]{inputenc}
\usepackage[T1]{fontenc}
\usepackage{subcaption}
\usepackage{multirow}
\usepackage{algorithmic}
\usepackage{algorithm}
\usepackage{optidef}

\title{Parallelizable Search-Space Decomposition for Large-Scale Combinatorial Optimization Problems Using Ising Machines}

\author[1,2]{Eiji Kawase}
\author[2,3]{Shuta Kikuchi}
\author[1]{Hideaki Tamai}
\author[2,3,4,5,*]{Shu Tanaka}
\affil[1]{Oki Electric Industry Co.,Ltd., Saitama, 335-8510, Japan}
\affil[2]{Graduate School of Science and Technology, Keio University, Kanagawa, 223-8522, Japan}
\affil[3]{Keio University Sustainable Quantum Artificial Intelligence Center (KSQAIC), Keio University, Tokyo, 108-8345, Japan}
\affil[4]{Department of Applied Physics and Physico-Informatics, Keio University, Kanagawa, 223-8522, Japan}
\affil[5]{Human Biology-Microbiome-Quantum Research Center (WPI-Bio2Q), Keio University, Tokyo, 108-8345, Japan}

\affil[*]{shu.tanaka@keio.jp}


\begin{abstract}
Combinatorial optimization problems are crucial in industry. 
However, many COPs are NP-hard, causing the search space to grow exponentially with problem size and rendering large-scale instances computationally intractable.
Conventional solvers typically treat problems as monolithic entities, leading to significant efficiency degradation as structural complexity increases. 
To address this issue, we propose a novel search-space decomposition method that leverages the inherent structure of variables to systematically reduce the size of the master problem.
We formulate interaction costs between variables and individual variable costs as a constrained maximum cut problem and convert it into a quadratic unconstrained binary optimization formulation using penalty terms.
An Ising-model solver is used to rapidly decompose the problem into independent small-scale subproblems, which are subsequently solved in parallel using mathematical optimization solvers.
We validated this method on the capacitated vehicle routing problem.
Results demonstrate three significant benefits: a substantial enhancement in feasible solution rates, accelerated convergence, achieving in 1 min the accuracy that the naive method required 30 min to reach, and a variable reduction of up to 95.32\%.
These findings suggest that search-space decomposition is a promising strategy for efficiently solving large-scale combinatorial optimization problems.
\end{abstract}

\begin{document}

\flushbottom
\maketitle

\thispagestyle{empty}

\section*{Introduction}
Combinatorial optimization problems (COPs) are defined as problems of finding a combination of discrete variables that minimize or maximize an objective function under given constraints~\cite{du1998handbook}.
These problems arise in diverse real-world fields, such as vehicle routing~\cite{dantzig1959truck}, production scheduling~\cite{pinedo2016scheduling}, facility location~\cite{daskin2013network}, and portfolio optimization~\cite{markowitz1952portfolio}, and play a critical role in industry.
However, many COPs are NP-hard, implying that the required computational effort may grow exponentially with problem size, making large-scale instances computationally challenging.
In addition, complex constraints significantly restrict the feasible region, further complicating the search process~\cite{garey1979computers}.
Exhaustively exploring such vast and complex search spaces to guarantee optimality is computationally impractical for most realistic large-scale instances.
To address this intractability, various approaches have been developed.
From a mathematical perspective, approximation algorithms~\cite{williamson2011design} provide polynomial-time solutions with theoretical guarantees, such as Christofides' algorithm~\cite{christofides1976worst} for the traveling salesman problem (TSP).
However, deriving such guarantees for complex, highly constrained industrial problems is often infeasible.
Consequently, metaheuristics such as simulated annealing~\cite{kirkpatrick1983optimization}, tabu search~\cite{glover1986future}, genetic algorithms~\cite{holland1975adaptation}, particle swarm optimization~\cite{kennedy1995particle}, and ant colony optimization~\cite{dorigo1996ant}, have become standard approaches for obtaining high-quality approximate solutions within a reasonable time frame~\cite{gendreau2019handbook}.
In parallel, commercial mathematical optimization solvers (e.g., Gurobi~\cite{gurobi}, CPLEX~\cite{cplex}) have evolved to incorporate sophisticated heuristics alongside exact methods such as branch-and-bound.
Despite these advancements, conventional solvers typically treat the problem as a single monolithic task.
Consequently, as the problem size increases, the search space expands rapidly and structural parallelization becomes challenging.
As a result, large-scale problems with complex structures may require prohibitively long computation times to obtain practically satisfactory solutions.

\begin{figure*}[htbp]
    \centering
    \includegraphics[width=.6\linewidth]{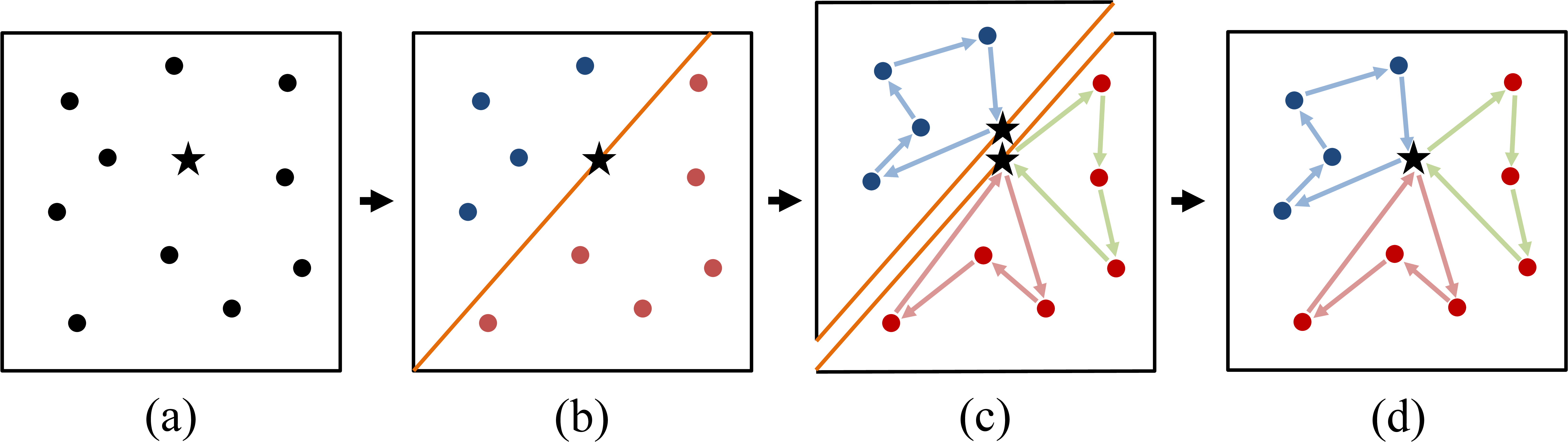}
    \caption{
    Overview of the proposed method. Circles represent customers and the star represents the depot. (a) Master problem. (b) Partitioning of variables into red and blue sets. (c) Parallel solution of subproblems using a mathematical optimization solver. (d) Integration of subproblem solutions. Reprinted with permission from~\cite{qce25_kawase}, ~\copyright 2025 IEEE.
    }
    \label{fig:proposal}
\end{figure*}

Decomposition methods provide a potential pathway to improved scalability by dividing large problems into manageable subproblems.
Classical techniques such as Benders decomposition~\cite{bnnobrs1962partitioning} or Dantzig--Wolfe decomposition~\cite{dantzig1960decomposition}, partition problems according to their mathematical structure, including separable constraints or block-angular formulations.
While theoretically sound, these methods generally require iterative coordination and information exchange between a master problem and subproblems, making fully independent parallel execution difficult.
To achieve high scalability through parallelism, a method is required that can rapidly partition a large-scale problem into multiple weakly coupled subproblems based on the intrinsic structure of the variables.
Our approach serves as a high-speed preprocessor that enhances the performance of conventional solvers, as shown in Fig.~\ref{fig:proposal}.

Therefore, this paper proposes a novel search-space decomposition method that exploits interaction costs between variables to generate weakly coupled subproblems.
The proposed method focuses on clusters of closely related variables inherent in the problem and decomposes it at points where relationships between variables are weak.
This strategy reduces the scale of the master problem by partitioning it into multiple weakly coupled subproblems, significantly improving search efficiency through parallel computing.
To realize this decomposition, we introduce the constrained maximum cut problem (CMC), which uses interaction costs between variables and individual variable costs as edge and node weights, respectively.
By casting variable partitioning of variables into a weighted graph cut framework with balance constraints, the search-space decomposition can be rigorously formulated as the CMC.
The maximum cut problem (MAX-CUT) is NP-hard, as shown by Karp~\cite{karp1972reducibility}.
Theoretical research has advanced in this area, including the approximation algorithm proposed by Goemans and Williamson~\cite{goemans1995improved}.
Additionally, MAX-CUT is equivalent to finding the ground state of an Ising spin glass model~\cite{barahona1982computational}.
Quantum annealing (QA) machines and other Ising machines are expected to efficiently search for low-energy states of the quadratic unconstrained binary optimization (QUBO) formulation or Ising model~\cite{tanaka2017quantum,tanahashi2019application,hauke2020perspectives,mohseni2022ising,chakrabarti2023quantum}.

An Ising model~\cite{Ising1925} is defined on a graph $G=(V,E)$, where $V$ and $E$ denote the sets of vertices and edges, respectively.
The Hamiltonian $H$ of the Ising model is defined as:
\begin{equation}
    \label{eq:hamiltonian}
    H=-\sum_{(i,j) \in E}J_{i,j}\sigma_{i}\sigma_{j}-\sum_{i \in V}h_{i}\sigma_{i},
\end{equation}
where $\sigma_{i} \in \lbrace +1, -1 \rbrace$ denotes the spin on vertex $i \in V$.
The coefficients on the right-hand side, $J_{i,j}$ and $h_{i}$ , denote the interaction strength on edge $(i, j) \in E$ and the external magnetic field at vertex $i \in V$, respectively. 
By incorporating the constraints of the CMC as penalty terms and expressing the resulting problem in a QUBO formulation~\cite{lucas2014ising} or its equivalent representation, the Ising model, rapid decomposition becomes possible using QA machines or Ising machines.
Once the search space is decomposed by Ising machines, the resulting subproblems are solved in parallel using mathematical optimization solvers.
The remainder of this paper is organized as follows: Results, Discussion, and Methods.
The Results section describes the proposed method and the experimental results. 
The Discussion section presents the insights gained from the experimental findings.
The Methods section describes the dataset used in the experiments, the parameter settings for the proposed method, and the procedures used to generate the graphs of the experimental results.

\section*{Results}
This section describes the proposed search-space decomposition method for large-scale COPs, the interaction costs used for decomposition, and the procedure for solving the resulting subproblems and constructing the final solution to the master problem.
Finally, we present experimental results evaluated in terms of the feasible solution rate (FS rate), convergence speed, and variable reduction rate (VR rate).

\subsubsection*{Formulation of the CVRP}
The CVRP addressed in this study is a variant of the vehicle routing problem (VRP) in which vehicles deliver customer demands from a single depot to multiple customers.
Specifically, it involves assigning vehicles to customers and determining the visiting order to minimize the total distance or total transportation cost.
In addition, each vehicle has a capacity limit, and all deliveries must satisfy these capacity constraints.

The CVRP is defined on a graph $G=(V,E)$, where $V=\{0,1, \dots,n\}$ denotes the set of vertices and $E$ denotes the set of edges. Vertex 0 represents the depot, and the set $S=V \setminus \{0\}$ denotes the customers.
Each customer $i \in S$ has a positive demand $d_{i}$.
Let $\mathcal{K} = \{1, \dots, K\}$ denote the set of vehicles.
There are $K$ vehicles located at the depot, each with capacity $Q$.
A non-negative travel cost $c_{i,j}$ is associated with each edge $(i,j) \in E$.
The binary variable $x_{i,j,k}$ takes the value 1 if vehicle $k$ travels from vertex $i$ to vertex $j$, and 0 otherwise.
In addition, the auxiliary variable $u_{i,k}$ represents the visiting order of customer $i$ on vehicle $k$'s route and is used for subtour-elimination constraints.
The formulation of the CVRP is as follows:
\begin{mini!}
    {x_{i,j,k}}{\sum_{i \in V} \sum_{j \in V, i \neq j} \sum_{k \in \mathcal{K}}c_{i,j}x_{i,j,k}}{}{\label{eq:obj}}
    \addConstraint{\sum_{j \in S} x_{0,j,k}}{= 1, \quad \forall k \in \mathcal{K} \label{eq:const1}}
    \addConstraint{\sum_{i \in S} x_{i,0,k}}{= 1, \quad \forall k \in \mathcal{K} \label{eq:const2}}
    \addConstraint{\sum_{j \in V, j \neq i} \sum_{k \in \mathcal{K}} x_{i,j,k}}{= 1, \quad \forall i \in S \label{eq:const3}}
    \addConstraint{\sum_{j \in V, j \neq i} \sum_{k \in \mathcal{K}} x_{j,i,k}}{= 1, \quad \forall i \in S \label{eq:const4}}
    \addConstraint{\sum_{j \in V, j \neq i} x_{j,i,k} - \sum_{j \in V, j \neq i} x_{i,j,k}}{= 0, \quad \forall i \in S, \forall k \in \mathcal{K} \label{eq:const5}}
    \addConstraint{\sum_{i \in S} d_{i} \sum_{j \in V, j \neq i} x_{i,j,k}}{\le Q, \quad \forall k \in \mathcal{K} \label{eq:const6}}
    \addConstraint{u_{i,k} - u_{j,k} + 1}{\le (|V|-1)(1 - x_{i,j,k}), \quad \forall i, j \in S, i \neq j, \forall k \in \mathcal{K} \label{eq:const7}}
    \addConstraint{0 \leq u_{i,k}}{\leq |S|, \quad \forall i \in S, \forall k \in \mathcal{K}. \label{eq:const8}}
\end{mini!}
Here, Eq.~\eqref{eq:obj} is the objective function, representing the minimization of the total travel cost.
In addition, the roles of the constraints are described as follows.
Equations~\eqref{eq:const1} and~\eqref{eq:const2} ensure that each vehicle $k$ departs from the depot and returns to the depot.
Equations~\eqref{eq:const3} and~\eqref{eq:const4} guarantee that exactly one vehicle visits and departs from each customer $i$.
Equation~\eqref{eq:const5} is the flow conservation constraint, indicating that a vehicle entering a customer must also leave that customer.
Equation~\eqref{eq:const6} is the capacity constraint, ensuring that the load of each vehicle does not exceed $Q$.
Equations~\eqref{eq:const7} and~\eqref{eq:const8} eliminate subtours, defined as loops that do not pass through the depot.

In this study, we use the capacity utilization rate (CUR) as an indicator of constraint tightness.
The CUR is defined as the ratio of the total demand to the total vehicle capacity:
\begin{equation}
    \label{eq:cur}
    \mathrm{CUR}=\frac{\sum_{i \in S}d_{i}}{KQ}.
\end{equation}
A higher CUR indicates tighter capacity constraints and requires more accurate decomposition.

\subsection*{Variable-based search-space decomposition algorithm}
This section describes Algorithm~\ref{alg:partition}, which decomposes the search space.
As a preprocessing step, we identify the index set to be partitioned for search-space decomposition.
Typically, an index set with large cardinality is selected because its partitioning substantially reduces the subproblem size.
Examples include the customer set ($S = V \setminus \{0\}$) in the VRP and the item set in the multiple knapsack problem.
We then set an upper bound $\mathrm{MAX}_\mathrm{variables}$ on the subproblem size to ensure that each resulting subproblem remains manageable for a mathematical optimization solver.
The algorithm stores the initial set in a list $\mathcal{L}$ and repeats the following procedure until $\mathcal{L}$ becomes empty.
At each iteration, the algorithm retrieves a set $S_{\mathrm{current}}$ from $\mathcal{L}$.
If $|S_{\mathrm{current}}| \leq \mathrm{MAX}_\mathrm{variables}$, it is added to the final list of subsets $S_{\mathrm{final}}$.
Otherwise, $S_{\mathrm{current}}$ is decomposed into two subsets $S_1$ and $S_2$ using the CMC, and these subsets are returned to $\mathcal{L}$.
By recursively repeating these steps, the search space is decomposed until all sets eventually satisfy the specified upper bound.

\begin{algorithm}[htbp]
    \caption{Recursive search-space decomposition}
    \label{alg:partition}
    \begin{algorithmic}[1]
    \REQUIRE Initial set of variables $S_{\mathrm{initial}}$, Size threshold $\mathrm{MAX}_\mathrm{variables}$
    \ENSURE A set of subproblems variables $S_{\mathrm{final}}$, where for each $s \in S_{\mathrm{final}}$, $|s| \leq \mathrm{MAX}_\mathrm{variables}$
    \STATE $\mathcal{L} \leftarrow \emptyset$
    \STATE $S_{\mathrm{final}} \leftarrow \emptyset$
    \STATE Add $S_{\mathrm{initial}}$ to $\mathcal{L}$
    \WHILE{$\mathcal{L}$ is not empty}
        \STATE $S_{\mathrm{current}} \leftarrow \text{Remove from } \mathcal{L}$
        \IF{$|S_{\mathrm{current}}| > \mathrm{MAX}_\mathrm{variables}$}
            \STATE $(S_1, S_2) \leftarrow \mathrm{CMC}(S_{\mathrm{current}})$
            \STATE Add $S_1$ to $\mathcal{L}$
            \STATE Add $S_2$ to $\mathcal{L}$
        \ELSE
            \STATE $S_{\mathrm{final}} \leftarrow S_{\mathrm{final}} \cup \{S_{\mathrm{current}}\}$
        \ENDIF
    \ENDWHILE
    \RETURN $S_{\mathrm{final}}$
    \end{algorithmic}
\end{algorithm}

\subsection*{Decomposition of search spaces using the CMC}
The CMC-based decomposition algorithm for the search space was first proposed in our previous work~\cite{qce25_kawase}.
This section explains the CMC formulation used to decompose the search space in Algorithm~\ref{alg:partition}.
The CMC is formulated using the binary decision variable $x_{i}$ in the objective function Eq.~\eqref{eq:MAX-CUT} and the constraint Eq.~\eqref{eq:cmc_const} as follows:
\begin{maxi!}
    {x_{i}}{\sum_{(i,j) \in E}W_{i,j}\left(-2x_{i}x_{j}+x_{i}+x_{j} \right)}{}{\label{eq:MAX-CUT}}
    \addConstraint{\sum_{i \in S}h_{i}\left(x_{i}-\alpha\right)}{= 0. \label{eq:cmc_const}}
\end{maxi!}
Here, by utilizing the interaction costs between variables $W_{i,j}$ as edge weights and the individual variable costs $h_i$ as node weights, the multivariate problem is partitioned into two weakly coupled sets of variables.
The parameter $\alpha$ ($0 \leq \alpha \leq 1$) controls the desired fraction of the total individual variable cost assigned to one of the two subsets.
Furthermore, this formulation corresponds to a quadratic binary optimization problem derived from MAX-CUT.
The objective in Eq.~\eqref{eq:MAX-CUT} corresponds to the weighted cut value of a graph, where assigning $x_i \in \{0,1\}$ determines the partition of the vertex set.
By incorporating the constraint as a quadratic penalty term into the objective function, the problem can be reformulated as a QUBO formulation as follows:
\begin{equation}
    \label{eq:CMC}
    H_{\textrm{CMC}}=\sum_{(i,j) \in E}W_{i,j}\left(2x_{i}x_{j}-x_{i}-x_{j} \right) + \mu \left(\sum_{i}h_{i}\left(x_{i}-\alpha\right)\right)^2.
\end{equation}
Here, $\mu$ is a hyperparameter that controls the strength of the constraint term.
Note that the sign change between Eqs.~\eqref{eq:MAX-CUT} and~\eqref{eq:CMC} reflects the conversion from a maximization problem to an equivalent energy minimization formulation.
In Eq.~\eqref{eq:CMC}, the decomposition is achieved by maximizing the interaction costs between variables while enforcing balance between the two subsets through the parameters $\mu$ and $\alpha$.

\subsection*{Application of the proposed method to the CVRP}
This section describes how the search-space decomposition based on Eq.~\eqref{eq:CMC} is applied to large-scale CVRP instances.
In the CVRP, the size of the search space in the CVRP primarily depends on the number of customers and grows exponentially as this number of customers increases.
Therefore, the method focuses on partitioning the customer index set.
In Eq.~\eqref{eq:CMC}, the travel cost between customers is used as the edge weight $W_{i,j}$ representing interaction costs, and the customer demand is used as the node weight $h_{i}$, enabling effective partitioning of the customer set.
Furthermore, this paper examines two definitions of the interaction cost $W_{i,j}$ between customers.
\begin{itemize}
\item $D_{i,j}$: Distance-based decomposition
\item $T_{i,j}$: Angular-based decomposition
\end{itemize}

\subsubsection*{Distance-based decomposition}
\begin{figure}[htbp]
    \centering
    \begin{minipage}[b]{.3\columnwidth}
        \centering
        \includegraphics[width=.95\columnwidth]{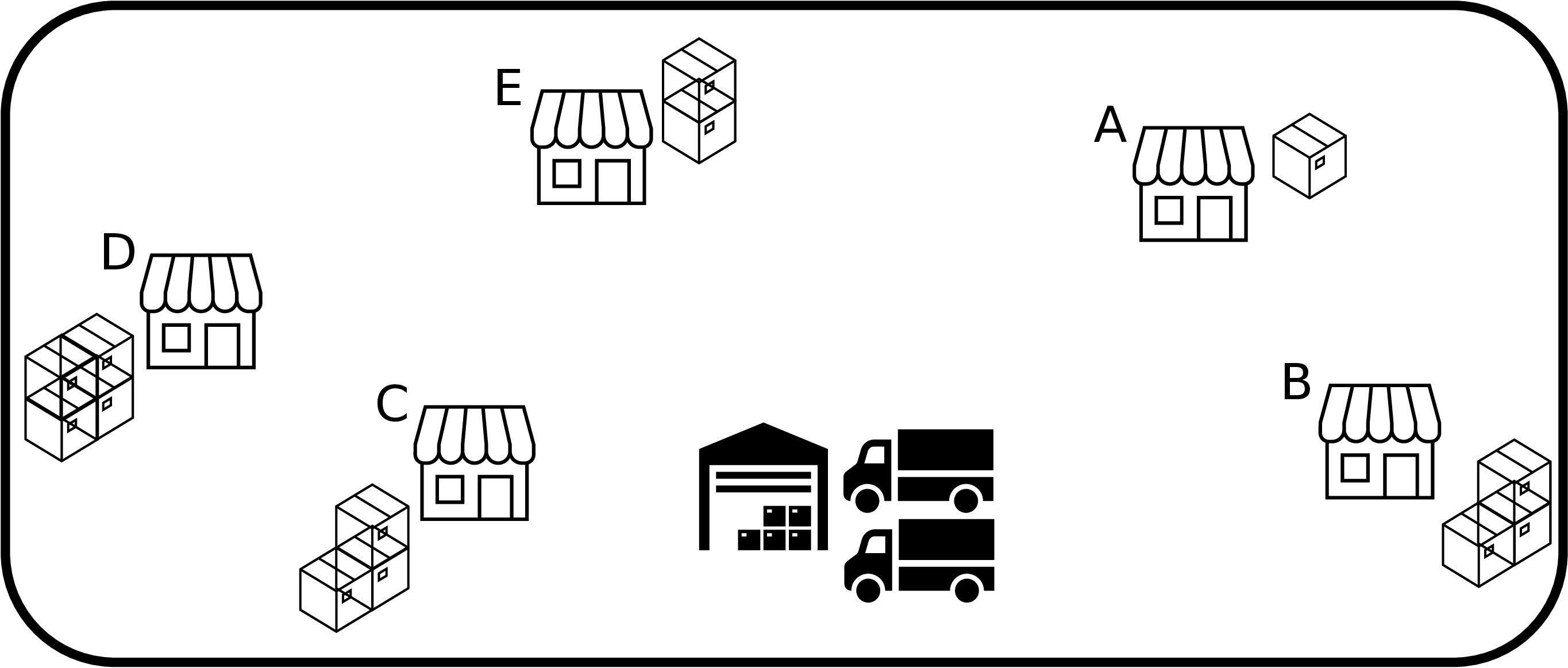}
        \subcaption{}
        \label{fig:pm1-a}
    \end{minipage}
    \begin{minipage}[b]{.3\columnwidth}
        \centering
        \includegraphics[width=.95\columnwidth]{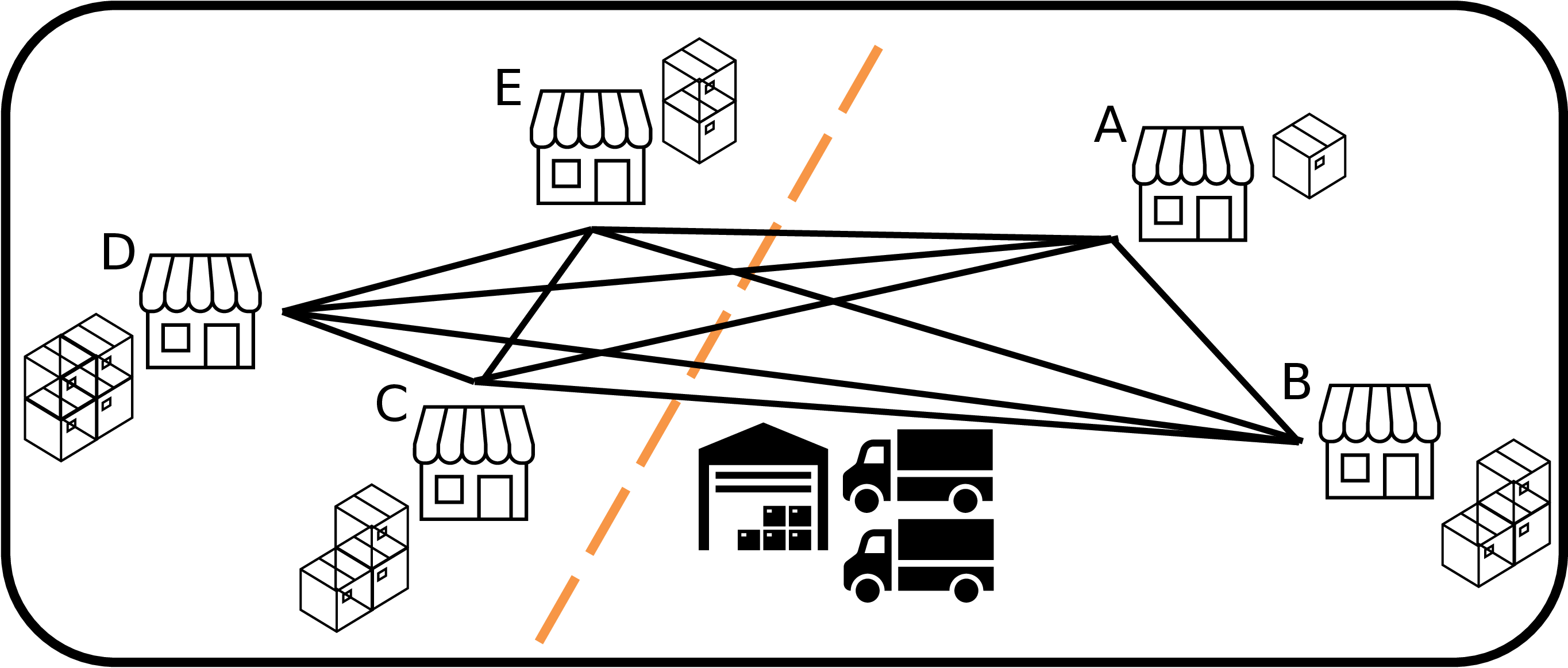}
        \subcaption{}
        \label{fig:pm1-b}
    \end{minipage}
    \begin{minipage}[b]{.3\columnwidth}
        \centering
        \includegraphics[width=.95\columnwidth]{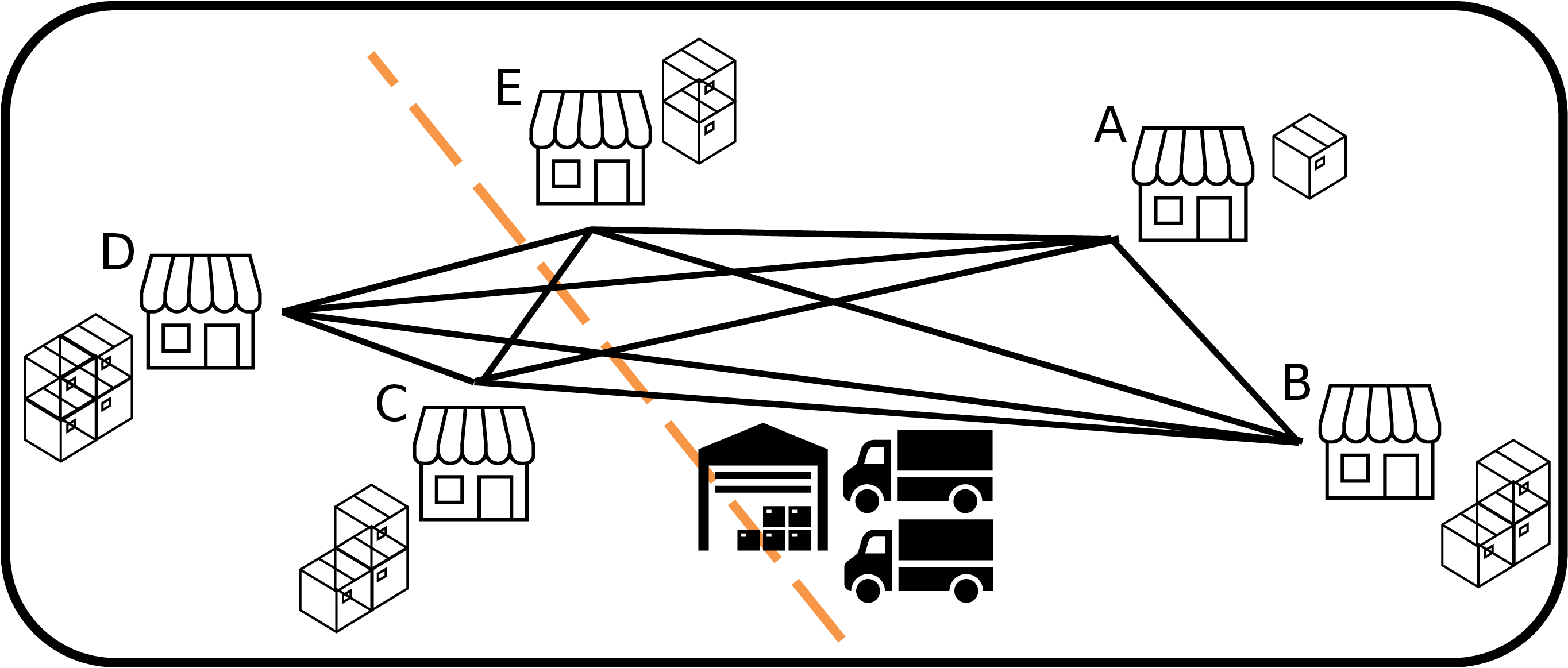}
        \subcaption{}
        \label{fig:pm1-c}
    \end{minipage}\\
    \caption{
    Illustration of DBD. (a) Delivery area of the master problem. (b) Decomposition result with $\mu = 0$ and $\alpha = 0.5$. (c) Decomposition result with $\mu > 0$ and $\alpha = 0.5$. The orange dashed line indicates the edges that are cut. Home icons (A–E) denote customers, the warehouse icon denotes the depot, vehicle icons indicate the number of vehicles, and cardboard box icons indicate customer demands. In this example, the demands of customers A, B, C, D, and E are 1, 3, 3, 4, and 2, respectively.
    }
    \label{fig:proposed_methods_dij}
\end{figure}

The first method employs a distance metric based on physical proximity between customers as the interaction cost in the CMC.
We refer to this method as distance-based decomposition (DBD).
The interaction cost between variables, denoted as $D_{i,j}$, is defined as the Euclidean distance between customers $i$ and $j$.
Because the underlying MAX-CUT objective seeks to maximize the weighted cut value, larger distances encourage edges between distant customers to be cut, placing them in different subsets, whereas nearby customers tend to remain in the same subset.
This decomposition method is expected to reduce unnecessary long-distance travel when constructing the final delivery route.
Furthermore, the demand quantity for each customer is incorporated as the node weight in the formulation.
This design is motivated by the potential demand imbalance may arise between the decomposed subsets.
If the total demand becomes skewed within each set, solving the CVRP for each set may yield solutions that require more vehicles than defined in the master problem.
Although such solutions are feasible at the subproblem level, they may violate the global vehicle constraints of the master problem.
The Hamiltonian $H_{\textrm{DBD}}$ for the QUBO formulation in the first method is defined as follows:
\begin{gather}
    \label{eq:pm1}
    H_{\textrm{DBD}}=\sum_{(i,j) \in E}D_{i,j}\left( 2x_{i}x_{j}-x_{i}-x_{j} \right) + \mu \left(\sum_{i \in S}d_{i}\left(x_{i}-\alpha\right)\right)^2,
\end{gather}
where $D_{i,j}$ represents the distance between customers $i$ and $j$, and $d_{i}$ indicates the demand quantity for customer $i$.
Furthermore, the demand balance coefficient $0 \leq \alpha \leq 1$ represents the desired proportion of total demand assigned to one subset when vehicle capacities are identical. The coefficient $\alpha$ is defined as follows:
\begin{gather}
    \label{eq:alpha}
    \alpha = \frac{\lfloor K / 2 \rfloor}{K}.
\end{gather}
The demand balance coefficient $\alpha$ is determined based on the number of vehicles.
Therefore, the CMC formulation partitions customer variables by maximizing the total weighted cut value based on $D_{i,j}$ while enforcing demand balance through the parameters $\mu$ and $\alpha$.
Figure~\ref{fig:pm1-a} illustrates the delivery area of the master problem.
Figure~\ref{fig:pm1-b} illustrates the decomposition performed using DBD when there is no constraint term (i.e., $\mu=0$), while Fig.~\ref{fig:pm1-c} illustrates the case with a constraint term (i.e., $\mu>0$).

Here, there are two vehicles, and $\alpha=0.5$ is obtained from Eq.~\eqref{eq:alpha}.
In Fig.~\ref{fig:pm1-b}, because the penalty term is absent, edges with larger distance weights are preferentially cut.
Consequently, the graph is partitioned into two sets: $\{A, B\}$ and $\{C, D, E\}$.
The total demand in each set is $\sum_{i \in \{A, B\}}d_{i}=4$ and $\sum_{i \in \{C, D, E\}}d_{i}=9$, respectively.
In Fig.~\ref{fig:pm1-c}, the penalty term is applied to balance the demand between the two subsets while still favoring the cutting of edges with large distance weights.
Consequently, the graph is partitioned into $\{A, B, E\}$ and $\{C, D\}$.
This example illustrates the trade-off between maximizing the weighted cut value and satisfying the demand balance constraint controlled by $\mu$.

\subsubsection*{Angular-based decomposition}
\begin{figure*}[htbp]
    \centering
    \includegraphics[width=.35\columnwidth]{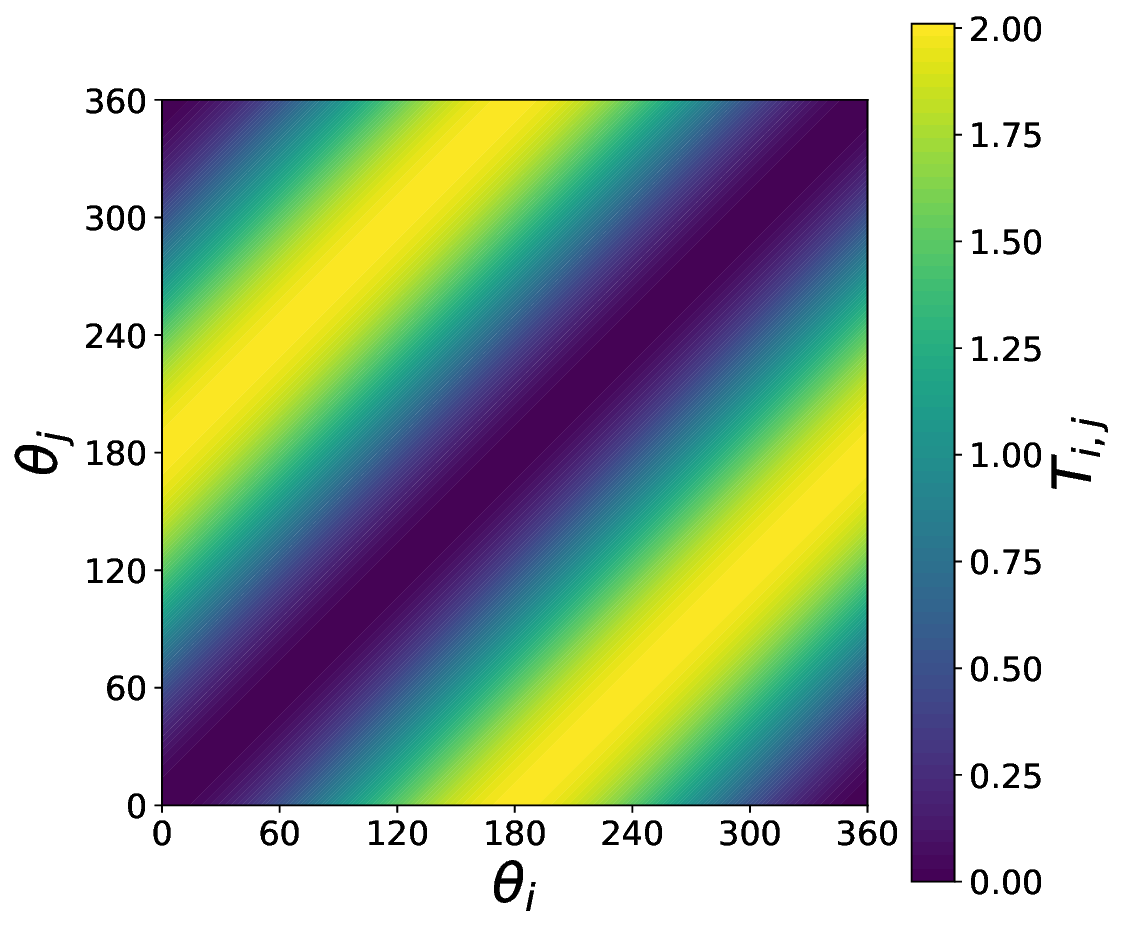}
    \caption{
    Values of the dissimilarity measure $T_{i,j}$ as a function of $\theta_i$ and $\theta_j$.
    }
    \label{fig:theta_ij}
\end{figure*}

The second method employs an angular indicator based on the relative direction from the depot.
We refer to this method as angular-based decomposition (ABD).
First, the location coordinates of the depot and customers in Euclidean space are transformed into a polar coordinate system centered at the depot.
Then, the dissimilarity measure derived from the angular difference between two customers is defined as a feature representing their relative angular positions with respect to the depot, denoted as $T_{i,j}$.
The dissimilarity value $T_{i,j}$ is defined using $\theta$ as follows:
\begin{gather}
    \label{eq:theta_ij}
    T_{i,j}=1-\cos{\left(\theta_{i}-\theta_{j}\right)},
\end{gather}
where $\theta_i$ and $\theta_j$ denote the polar angles of customers $i$ and $j$ measured from the depot.
Figure~\ref{fig:theta_ij} illustrates the values of Eq.~\eqref{eq:theta_ij} as a function of $\theta_i$ and $\theta_j$.

\begin{figure}[htbp]
    \centering
    \begin{minipage}[b]{.49\columnwidth}
        \centering
        \includegraphics[width=.5\columnwidth]{./fig/pm2-1.eps}
        \subcaption{}
    \end{minipage}
    \begin{minipage}[b]{.49\columnwidth}
        \centering
        \includegraphics[width=.5\columnwidth]{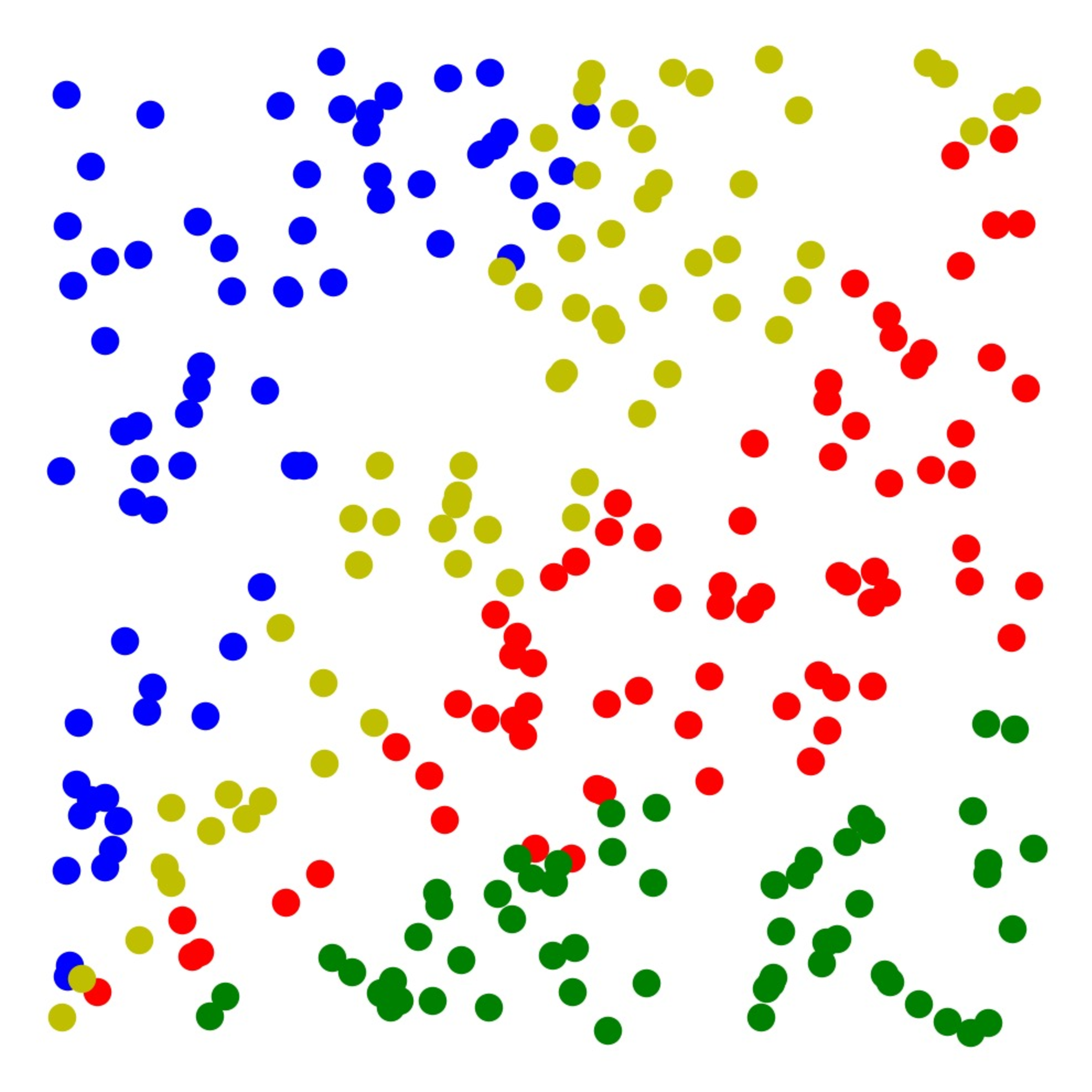}
        \subcaption{}
    \end{minipage}\\
    \caption{Schematic illustration of ABD. (a) Mapping of all customers onto a polar coordinate system with the depot as the origin. (b) Example result of solving the CMC using $T_{i,j}$. The circles and the star denote customers and the depot, respectively.}
    \label{fig:tij_image}
\end{figure}

Using this feature enables decomposition of customer variables while explicitly incorporating their positional relationship with the depot, which was not considered in the first method.
Specifically, the larger the angular separation between two customers relative to the depot, the more likely they are to be classified into different clusters; conversely, smaller angular differences favor assignment to the same subset.
As a result, customers located in similar directions from the depot are aggregated into the same cluster.
This radial grouping reflects the geometric structure of the service area and facilitates construction of more efficient delivery routes in the subsequent routing phase.
Unlike DBD, ABD captures directional structure relative to the depot, which is particularly relevant in capacitated routing problems where route efficiency depends on radial customer distribution.
The Hamiltonian $H_{\textrm{ABD}}$ for the QUBO formulation of ABD is defined as follows:
\begin{gather}
    \label{eq:pm2}
    H_{\textrm{ABD}}=\sum_{(i,j) \in E}T_{i,j}\left( 2x_{i}x_{j}-x_{i}-x_{j} \right) + \mu \left(\sum_{i \in S}d_{i}\left(x_{i}-\alpha\right)\right)^2,
\end{gather}
where $T_{i,j}$ represents the dissimilarity value defined in
Eq.~\eqref{eq:theta_ij}, and $d_i$ and $\alpha$ are as defined in the previous section.
Figure~\ref{fig:tij_image} illustrates the decomposition performed using ABD.
Here, each customer's color represents the group to which the customer belongs, and the decomposition is performed such that customers located in the same angular direction from the depot belong to the same set.

\subsubsection*{Subproblems after search-space decomposition}
After recursively applying the CMC-based decomposition until the stopping criteria are satisfied, the customer set of the master problem is partitioned into multiple subsets.
The number of vehicles assigned to each subproblem is determined based on the total demand of the subset $S_\mathrm{sub}$ as 
$K_\mathrm{sub}=\left\lceil ({\sum_{i \in S_\mathrm{sub}} d_i})/{Q} \right\rceil$,
where $Q$ denotes the vehicle capacity.
Note that, due to the ceiling operation in the definition of $K_{\mathrm{sub}}$, the total number of vehicles assigned to all subproblems, $\sum_{\mathrm{sub}} K_{\mathrm{sub}}$, may exceed the number of vehicles $K$ in the master problem. 
However, the demand balance constraint controlled by $\alpha$ in the CMC formulation promotes an approximately even distribution of total demand across subsets, thereby limiting excessive fragmentation of vehicle assignments. 
Each CVRP subproblem is constructed by adding the depot and the corresponding number of vehicles to the customer subset.
The routes obtained from all subproblems are then integrated to construct the solution to the master problem.
The final solution is defined as the union of all routes generated for each subset, without additional re-optimization across subsets. 

\subsection*{FS rate and solution accuracy}
We calculated the FS rate, average gap (Avg. best-known solution (BKS) gap), and minimum gap (Min. BKS gap) with respect to the BKS for instances in which feasible solutions were obtained within 10 trials.
Let FS obj. and BKS obj. denote the objective values of a feasible solution and the BKS, respectively. 
The gap is defined as:
\begin{gather}
    \label{eq:gap}
    \rm{Gap}\,[\mathrm{\%}]=\frac{\rm{FS\ obj.}-\rm{BKS\ obj.}}{BKS\ obj.}\times100.
\end{gather}
The experimental results for each method are shown in Table~\ref{tab:result_main}.
As a baseline, hereafter referred to as the naive method, the CVRP is solved directly using only a mathematical optimization solver without decomposition.
As the mathematical optimization solver, we use the heuristic functionality of Gurobi (v12.0.0), a state-of-the-art commercial solver widely used in academia and industry.
For comparison purposes, the time limit for Gurobi was set to 30 min in the naive method. 
In the proposed method, each decomposed subproblem is solved independently, enabling parallel computation.
Therefore, the computation time was set to 30 min for each subproblem in the proposed method.
Consequently, the wall-clock time of the proposed method is at most 30 min, which is equal to that of the naive method.
In the proposed method, in addition to Gurobi, we used neal~\cite{neal}, an SA from the D-Wave Ocean SDK, to solve the CMC.
We report the computation time of SA, denoted as SA time, required for partitioning.
During decomposition, SA was executed 100 times per partitioning step, and the solution with the minimum energy was adopted.
The SA time increases with the number of variables because partitioning is performed iteratively.
The SA computation time is reported separately from the time limit imposed on  the subproblem optimization.

\begin{table*}[htbp]
    \begin{center}
        \centering
        \caption{Comparison of results: FS rate, BKS gap, and SA time}
        \label{tab:result_main}
        \begin{tabular}{cccccc}
            \hline \hline
            \textbf{Method} & \textbf{Instance} & \textbf{FS rate} & \textbf{Avg. BKS gap} & \textbf{Min. BKS gap} & \textbf{SA time} \\
            \hline
            \multirow{6}{*}{Naive method} 
             & M-n151-k12 & 100\% & 6.83\% & 6.83\% & - \\
             & M-n200-k17 & 100\% & 16.04\% & 15.81\% & - \\
             & X-n101-k25 & 0\% & - & - & - \\
             & X-n200-k36 & 0\% & - & - & - \\
             & X-n261-k13 & 100\% & 65.84\% & 64.35\% & - \\
             & X-n401-k29 & 0\% & - & - & - \\
             \hline
            \multirow{6}{*}{\shortstack{DBD}} 
             & M-n151-k12 & 100\% & 3.47\% & 3.47\% & 0.22s \\
             & M-n200-k17 & 100\% & 9.17\% & 9.17\% & 1.56s \\
             & X-n101-k25 & 0\% & - & - & - \\
             & X-n200-k36 & 30\% & 3.94\% & 3.39\% & 0.47s \\
             & X-n261-k13 & 100\% & 8.16\% & 8.16\% & 2.40s \\
             & X-n401-k29 & 10\% & 11.20\% & 11.20\% & 5.57s \\
             \hline
            \multirow{6}{*}{\shortstack{ABD}} 
             & M-n151-k12 & 100\% & 5.61\% & 5.61\% & 0.19s \\
             & M-n200-k17 & 100\% & 8.95\% & 8.95\% & 1.23s \\
             & X-n101-k25 & 0\% & - & - & - \\
             & X-n200-k36 & 60\% & 5.61\% & 4.59\% & 0.48s \\
             & X-n261-k13 & 100\% & 9.49\% & 9.01\% & 2.14s \\
             & X-n401-k29 & 100\% & 8.66\% & 6.92\% & 5.41s \\
             \hline
        \end{tabular}
    \end{center}
\end{table*}

The results indicate that for large-scale instances with a CUR exceeding approximately 98\% (e.g., X-n200-k36, X-n401-k29, listed in Table~\ref{tab:CVRPLIB_info}), the naive method failed to find feasible solutions within the time limit.
In contrast, the proposed method successfully obtained feasible solutions for these instances.
Notably, ABD achieved a 100\% success rate for the X-n401-k29 instance.
Regarding solution accuracy, the proposed method demonstrated substantial improvements over the naive method in terms of the gap to the BKS (Table~\ref{tab:result_main}).
In instances where the naive method found feasible solutions, the proposed method improved solution accuracy by approximately 3--50 percentage points.
For example, in the X-n261-k13 instance, the naive method had an average BKS gap of 65.84\%, whereas the proposed method using DBD reduced this value to 8.16\%.
Notably, for all instances where feasible solutions were obtained, ABD consistently provided high-quality solutions, with the average BKS gap ranging from 5.61\% to 9.49\%.

\subsection*{Improved convergence speed of solutions}
\begin{figure}[htbp]
    \begin{minipage}[t]{0.49\linewidth}
        \centering
        \includegraphics[width=0.78\linewidth]{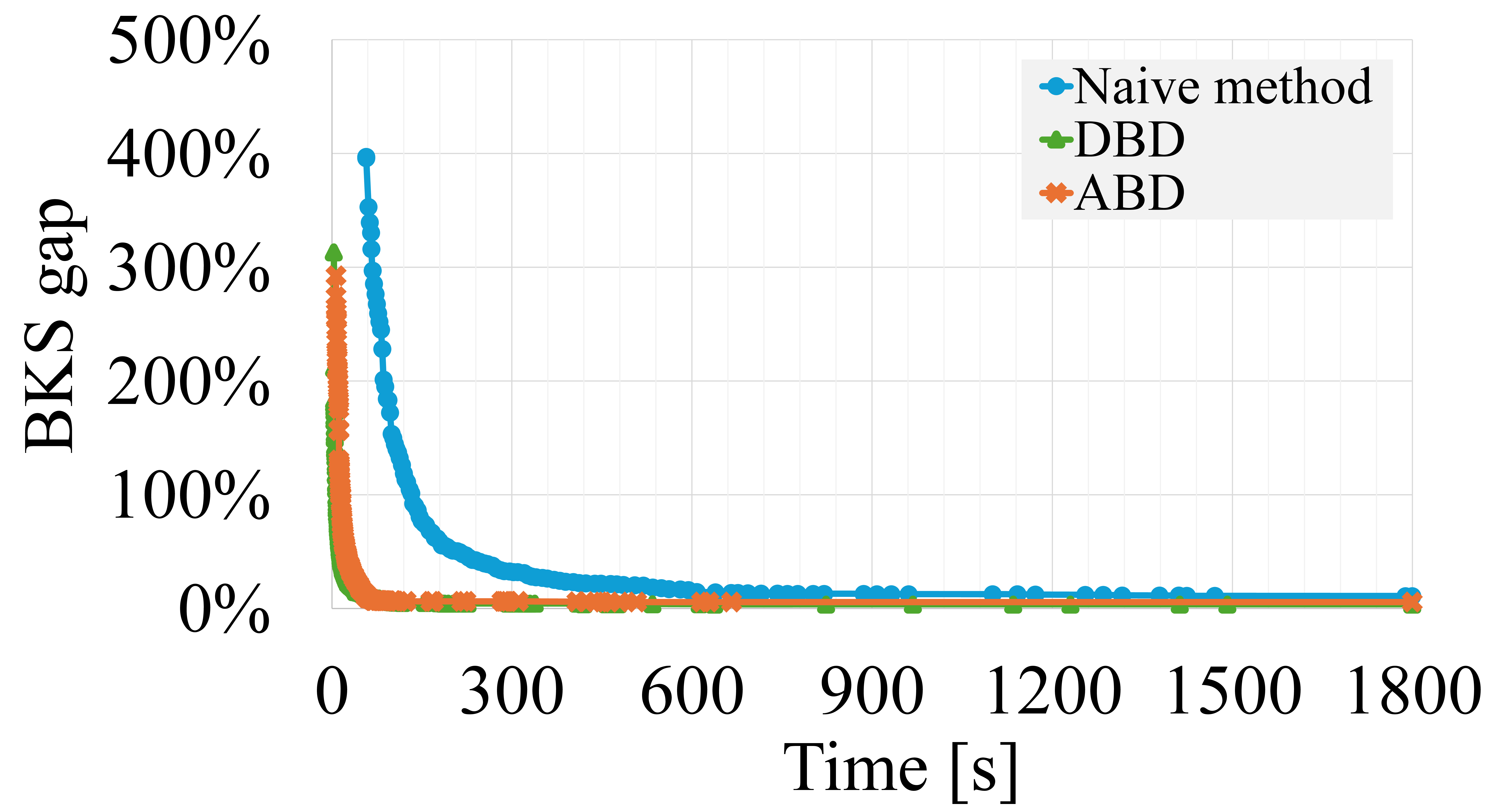}
        \subcaption{}
        \label{M-n151-k12}
    \end{minipage}
    \begin{minipage}[t]{0.49\linewidth}
        \centering
        \includegraphics[width=0.78\linewidth]{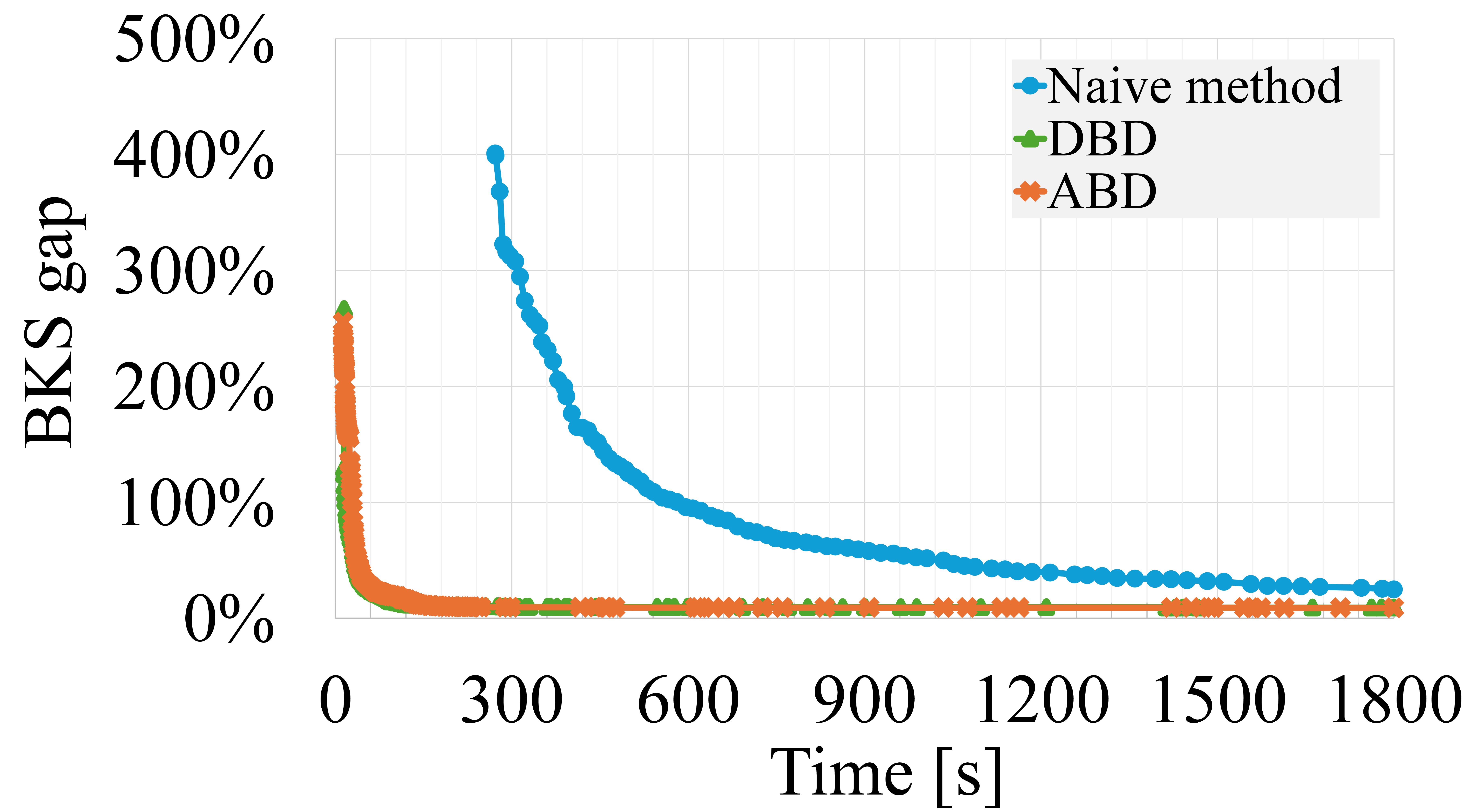}
        \subcaption{}
        \label{M-n200-k17}
    \end{minipage}\\
    \begin{minipage}[t]{1.0\linewidth}
        \centering
        \includegraphics[width=0.39\linewidth]{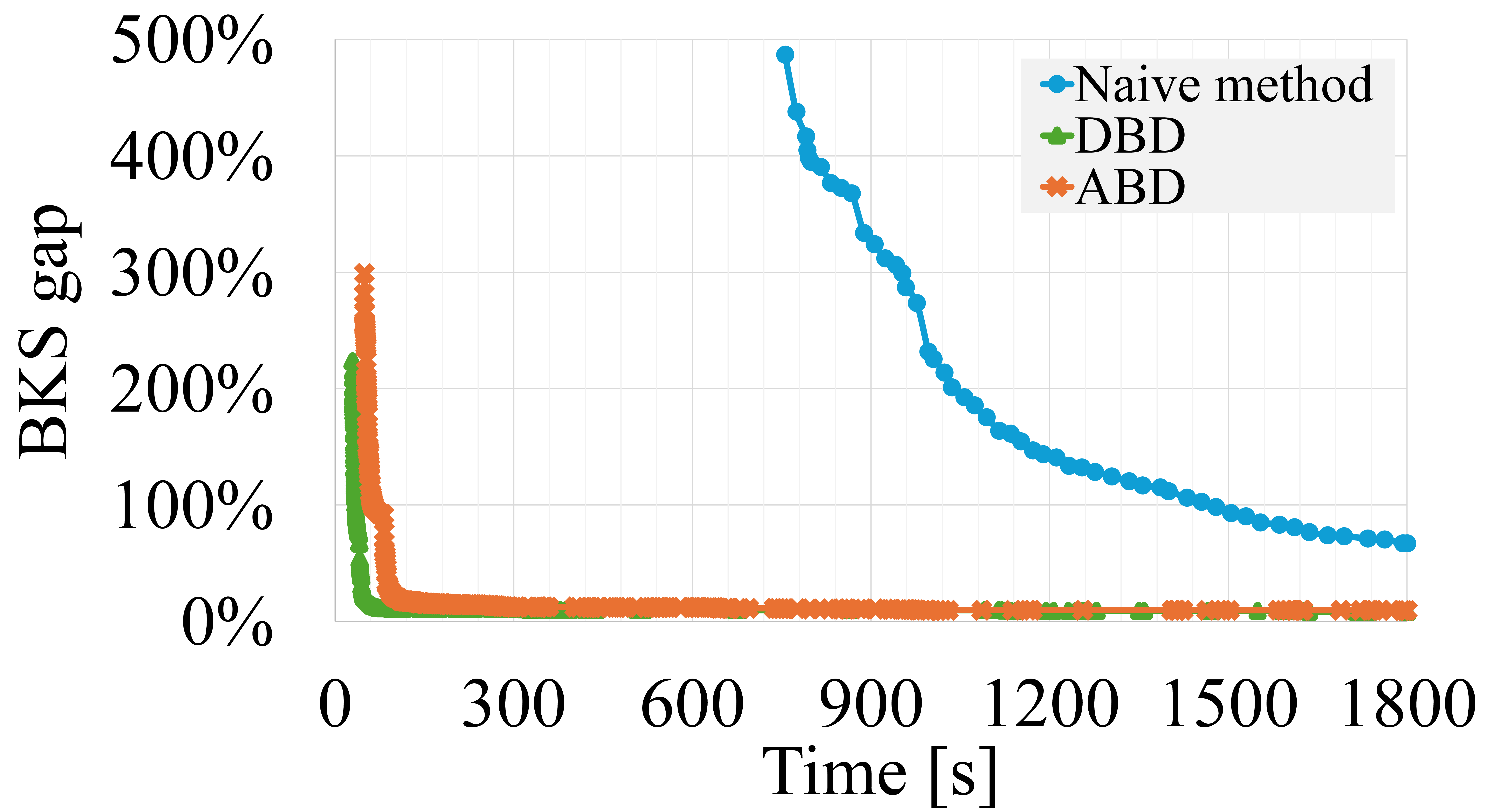}
        \subcaption{}
        \label{X-n261-k13}
    \end{minipage}
    \caption{
    Convergence performance over time for the naive method and the two proposed methods. (a) M-n151-k12. (b) M-n200-k17. (c) X-n261-k13.
    }
    \label{fig:result_graph}
\end{figure}

We measured the time from acquisition of the initial feasible solution until the gap with respect to the BKS stabilized.
Figure~\ref{fig:result_graph} shows the convergence curves of the gap for the three instances in which feasible solutions were obtained using all methods.
The proposed method converged substantially faster than the naive method. Specifically, the solution accuracy achieved by the naive method in 30 min was attained by the proposed method in approximately 1 min.
Furthermore, the advantage in convergence speed became more pronounced as problem size increased.

\subsection*{Variable reduction through search-space decomposition}
Table~\ref{tab:result_variables} presents the difference in the number of variables as an indicator of the search space reduction achieved by the proposed method.
The number of variables is calculated as $N_\mathrm{variables}=\left(|V||S|K+|S|K\right)$, where $|V|$ denotes the number of vertices including both customers and the depot, $|S|$ denotes the number of customers, and $K$ denotes the number of vehicles.
In the proposed method, the number of variables is calculated for each subproblem, and the sum is taken as the final number of variables.
The choice of interaction cost had no significant effect on the VR rate.
Compared with the naive method, the proposed method achieved reductions ranging from approximately 74.65\% to 95.32\%.
The reduction rate tended to increase as the number of customers increased, indicating scalability for large-scale problems.
These results quantitatively demonstrate that recursive decomposition substantially reduces the dimensionality of the optimization problem.

\begin{table*}[htbp]
    \begin{center}
        \centering
        \caption{Number of variables and VR rates for the previous and proposed methods}
        \label{tab:result_variables}
        \begin{tabular}{cccccc}
            \hline \hline
            \multirow{2}{*}{\textbf{Instance}} & \multirow{1}{*}{\textbf{Naive method}} & \multicolumn{2}{c}{\textbf{DBD}} & \multicolumn{2}{c}{\textbf{ABD}}\\ \cline{3-4}\cline{5-6}
            & $N_\mathrm{variables}$ & $N_\mathrm{variables}$ & VR rate & $N_\mathrm{variables}$ & VR rate \\
            \hline
            M-n151-k12 & 273,600 & 69,300 & 74.67\% & 69,348 & 74.65\% \\
            M-n200-k17 & 679,983 & 99,952 & 85.30\% & 101,896 & 85.01\% \\
            X-n200-k36 & 1,439,964 & 224,235 & 84.43\% & 221,796 & 84.60\% \\
            X-n261-k13 & 885,560 & 57,438 & 93.51\% & 58,334 &  93.41\% \\
            X-n401-k29 & 4,663,200 & 218,613 & 95.31\% & 218,370 & 95.32\% \\
            \hline
        \end{tabular}
  \end{center}
\end{table*}

\section*{Discussion}
The results of this study indicate that recursive search-space decomposition using the CMC, solved by Ising machines, is a highly effective hybrid approach for large-scale COPs.
While conventional solvers struggle with monolithic instances because of exponential expansion of the search space, the proposed method systematically reduces complexity by partitioning the master problem into independent subproblems.
A key finding is the clear advantage of ABD over DBD in high-CUR instances. By utilizing the dissimilarity value $T_{i,j}$ based on the relative direction from the depot, ABD tends to produce subproblems that include a more balanced distribution of customers from both near and distant locations. 
This approach facilitates generation of more efficient delivery routes compared with DBD, which primarily emphasizes physical proximity without explicitly incorporating directional structure relative to the depot.
However, for the X-n101-k25 instance, which has the highest CUR (0.9994) among all tested instances, none of the methods, including ABD, found feasible solutions. 
This finding suggests that when the capacity constraint is nearly saturated, even balanced decomposition cannot guarantee feasibility, and additional strategies, such as relaxing the vehicle count or incorporating slack variables, may be required.

Another practical consideration is tuning of the penalty coefficient $\mu$. In the current implementation, $\mu$ is incremented in steps whose granularity depends on the scale of the interaction cost, with integer steps for DBD and finer steps of 0.001 for ABD.
This instance-dependent tuning procedure may limit out-of-the-box applicability of the method.
Developing an adaptive or automated strategy for setting $\mu$, for example, based on the ratio of objective and penalty scales, is an important direction for future work.

It is worth noting that classical graph partitioning tools such as METIS~\cite{karypis1998fast} and clustering methods such as $k$-means can also partition customer sets. 
However, these methods do not directly optimize the weighted cut value under a demand balance constraint, which is the key mechanism by which the proposed method preserves problem structure during decomposition. 
A systematic comparison with such alternative partitioning strategies remains an important topic for future investigation.

The computational overhead of this decomposition is minimal, with SA times on the order of seconds.
This high-speed preprocessing enables mathematical optimization solvers to reach high-quality solutions up to 30 times faster than the naive method in certain instances.
Although the experiments were conducted on a single multi-core machine (Apple M3 Max), the proposed method is inherently suitable for distributed computing environments, where each subproblem can be dispatched to a separate node. 
In such settings, the wall-clock time advantage is expected to be even more pronounced, particularly for large-scale industrial
instances that yield many subproblems.
Furthermore, a VR rate of up to 95.32\% indicates the scalability of this approach for real-world industrial applications.
Future work will explore the application of this framework to other optimization domains such as scheduling and facility location problems.

\section*{Methods}
This section describes the experimental setup used to compare the naive and proposed methods.

\subsection*{Experimental setup}
We evaluated the proposed method using benchmark instances from CVRPLIB~\cite{cvrplib} for the CVRP, a representative COP.
Table~\ref{tab:CVRPLIB_info} summarizes the characteristics of the six instances used in this study.

\begin{table}[htbp]
  \begin{center}
    \centering
    \caption{CVRPLIB instance information}
    \label{tab:CVRPLIB_info}
    \begin{tabular}{cccccc}
        \hline \hline
        \textbf{Instance} & \textbf{The number of customers} & \textbf{The number of vehicles} & \textbf{Capacity} & \textbf{BKS} & \textbf{CUR} \\
        \hline
        M-n151-k12 & 150 & 12 & 200 & 1053 & 0.9313 \\
        M-n200-k17 & 199 & 17 & 200 & 1373 & 0.9371 \\
        X-n101-k25 & 100 & 25 & 206 & 27591 & 0.9994 \\
        X-n200-k36 & 199 & 36 & 402 & 58578 & 0.9856 \\
        X-n261-k13 & 260 & 13 & 1081 & 26558 & 0.9522 \\
        X-n401-k29 & 400 & 29 & 745 & 66154 & 0.9847 \\
        \hline
    \end{tabular}
  \end{center}
\end{table}

The instance names follow the standard CVRPLIB naming convention S-nY-kZ, where S denotes the instance set or series (e.g., M, X), nY represents the total number of nodes $Y$ including customers and the single depot and kZ indicates the number of vehicles $Z$ required for the BKS.

The solvers for both the naive and proposed methods were executed on a MacBook Pro equipped with an Apple M3 Max chip (16-core) and 128 GB of memory.

\subsection*{Details about solvers used in the experimental}
This section explains the solvers used to solve the CMC and CVRP in the naive and proposed methods.
As a baseline, hereafter referred to as the naive method, the CVRP is solved directly using only a mathematical optimization solver without decomposition.
As the mathematical optimization solver, we use the heuristic functionality of Gurobi (v12.0.0), a state-of-the-art commercial solver widely used in academia and industry.
For comparison purposes, the time limit for Gurobi was set to 30 min in the naive method. 
In the proposed method, each decomposed subproblem is solved independently, enabling parallel computation.
Therefore, the computation time was set to 30 min for each subproblem.
Consequently, the wall-clock time of the proposed method is at most 30 min, which is equal to that of the naive method.
In the proposed method, in addition to Gurobi, we used neal, an SA from the D-Wave Ocean SDK, to solve the CMC.
We report the computation time of SA, denoted as SA time, required for partitioning.
During decomposition, SA was executed 100 times per partitioning step, and the solution with the minimum energy was adopted.
The SA time increases with the number of variables because partitioning is performed iteratively.
The SA computation time is reported separately from the time limit imposed on subproblem optimization.
Experiments using Gurobi were conducted with default parameter settings except for the computation time.
Similarly, experiments using SA were conducted with default parameter settings except for the number of executions.

\subsection*{Parameter settings of proposed method}
This section explains the various parameters involved in solving the CMC in the proposed method.
First, the value of $\mathrm{MAX}_\mathrm{variables}$, which determines the recursive application of the CMC in Algorithm 1, was set to 100.
This choice was made because the Gurobi heuristic can compute an initial solution within tens of seconds at this scale.
The hyperparameter $\mu$, which represents the strength of the constraint term in the proposed method, was determined appropriately for each cost scale and each problem instance.
Specifically, in Eq.~\eqref{eq:pm1} for DBD, the range of $D_{i,j}$ is wide because it represents actual distances.
Therefore, $\mu$ was incremented as $\mu= \lbrace 0, 1, 2, 3,...\rbrace$ until the penalty term was sufficiently strong to enforce the demand balance specified by $\alpha$.
In contrast, in Eq.~\eqref{eq:pm2} for ABD, the range of $T_{i,j}$ is limited ($0 \leq T_{i,j} \leq 2$).
Therefore, $\mu$ was incremented in finer steps as $\mu = \{0, 0.001, 0.002, 0.003,\dots\}$.
The same stopping criterion as in DBD was applied.

\subsection*{Construction of convergence curves}
This section explains how the convergence curves in Fig.~\ref{fig:result_graph} are constructed to compare the naive method with the two proposed methods.
Because the proposed method solves each subproblem in parallel, the initial plot corresponds to the maximum time required to obtain initial feasible solutions across all subproblems and the gap between the resulting total objective value and the BKS.
Subsequently, whenever an improved solution is obtained for a subproblem, the elapsed wall-clock time is updated and the previous FS obj. is replaced with the FS obj. of the improved solution.
This procedure is repeated until the time limit is reached, yielding the convergence curves.

\section*{Data availability}
The data that support the findings of this study are available from the corresponding author upon reasonable request.

\bibliography{reference}

\section*{Acknowledgments}
This work was partially supported by the Council for Science, Technology, and Innovation (CSTI) through the Cross-ministerial Strategic Innovation Promotion Program (SIP), ``Promoting the application of advanced quantum technology platforms to social issues'' (Funding agency: QST), Japan Science and Technology Agency (JST) (Grant Number JPMJPF2221).
S.~Tanaka expresses gratitude to the World Premier International Research Center Initiative (WPI), MEXT, Japan, for their support of the Human Biology-Microbiome-Quantum Research Center (Bio2Q).
The authors would like to thank IEEE for permission to reuse Fig.~\ref{fig:proposal} from our previous conference paper~\cite{qce25_kawase}.We would like to thank Editage (www.editage.jp) for English language editing.

\section*{Author contributions}
E.~K. conceived the idea, performed simulations, and wrote the manuscript.
S.~K., H.~T., and S.~T. discussed and commented on the manuscript, and S.~T. supervised the project.
All authors reviewed the manuscript.

\end{document}